\documentstyle[aps,preprint,psfig]{revtex}
\tightenlines
\begin{document}

\title{{\hfill\normalsize NBI--97--33}\\
{\hfill\normalsize LBNL--40846}\\
Charmlets}

\author{J\"urgen Schaffner-Bielich}
\address{
Nuclear Science Division, Lawrence Berkeley National Laboratory,\\
University of California, Berkeley, CA 94720, USA}
\author{Axel P.~Vischer}
\address{
The Niels Bohr Institute, Blegdamsvej 17, DK-2100 Copenhagen \O, Denmark}
\maketitle

\begin{abstract}
We discuss the possibility, that multiply charmed exotics are produced
in ultrarelativistic heavy ion collisions. Production probabilities at
RHIC are estimated and found to be large enough to allow for the
possible detection of single and double charmed exotics. 
We demonstrate that charm matter is bound due to an attractive one-gluon
exchange potential. 
Furthermore, 
we study the color magnetic and color electric potential separately, 
as well as the group structure of such exotics and estimate their masses.
\end{abstract}

\section{Introduction}

Relativistic heavy ion collisions at very high energy 
will produce 
several charm quarks in one single event.
Estimates using perturbative QCD predict about 10 produced 
charm-anticharm quark pairs
at the Relativistic Ion Collider (RHIC) at Brookhaven
in one single central collision of two led nuclei at an bombarding energy of 
$\sqrt{s}= 200$ AGeV.
This offers the unique opportunity to study the possible existence of 
multiple charmed objects.

It is remarkable, that hadrons with two or more charm quarks, like the
$\Omega_{cc}$ and the $\Omega_{ccc}$ have not be seen so far (for a summary see
e.g. \cite{Richard94}). 
Certain exotic states with charm quarks have been discussed intensively in the
literature. Multiquark states have been studied in potential models using meson
exchange \cite{Mano93,Eric93,Torn94}.
The tetraquark with two heavy quarks of mass $m_Q$ and two light
(u,d,s) quarks is bound in the limit of $m_Q\to\infty$, but the situation 
for the
charm quark with $m_c\approx 1.5$ GeV is less clear
(see e.g.\ \cite{Carlson88} and references therein). 
A recent investigation finds that the heavy Tetraquark system is bound when
using pseudoscalar-meson exchange contributions \cite{Pepin97}.
The pentaquark with one anticharm quark and four light
quarks was proposed independently by Lipkin \cite{Lipkin87} and by Gignoux and
coworkers \cite{Gig87}. 
This state is bound by an
enhanced color magnetic binding energy. Hexaquark and Heptaquark 
states with charm have been
discussed also \cite{Chow95}.
There are proposals to search for the tetraquark and pentaquark in pp
collisions (for an overview see e.g.\ \cite{Moin96,MALL96}) at CHARM2000 and
CERN COMPASS.
Other exotics, especially for more than
two charm quarks, have not been considered so far.

Here, we investigate the structure of multiply charm objects
(charmlets), as well as their masses and stability.
First, estimates are given for the production rates of these objects
at RHIC using a simple coalescence approach. Then, the properties of charm bulk
matter are discussed. A modified bag model is introduced in section three
which is able to describe hadrons with and with out charm in the same approach
The color electric and magnetic interactions for charmlets are investigated in
section four. 
Finally, we give a summary and an outlook.


\section{Production of charmed exotics at RHIC}

The first and most basic question we have to address is, if there is any
chance to detect multi-quark exotics in heavy ion collisions. We
will try to answer this question in this section in the case of
$Au+Au$ collisions at RHIC. First we estimate
very crudely the production probability of multi--charmed hadrons and
then compare this production probability to detailed calculations within the
cascade model HIJING \cite{Lin97} using a coalescence code as
afterburner. An early estimate of heavy baryon production at RHIC and LHC
within a simple statistical model can
be found in \cite{LZ93}.

The first scale we have to determine is the total number of events
$N_{events}$ at
the RHIC collider in an average year. Every production rate or
probability $P$ has to be compared to this number. Only in
cases for which $ N_{events} P \gg 1$ we have a fair chance to measure
the exotic. Given the luminosity at RHIC 
$\approx 2\;10^{26}$ cm$^{-2}$ s$^{-1}$
and the geometric cross section of gold nuclei $\approx 3$ barn, we find
that there can be up to $2\;10^{10}$ heavy ion collision events per
year. This number has to be reduced by several factors. 
First, the
accelerator will run only part of the year.
In addition, the detectors
will only cover a small fraction of the full solid angle and
one has to face all kind of
inefficiencies. 
Finally, there is the branching ratio of the exotic state to measurable
(charged) final states.

Guided by the proposal for the experiments CHARM2000 and CERN COMPASS 
designed to measure charmed
exotics in pp collisions \cite{Moin96,MALL96},
one might be able to recover as little as $0.1\%$ of the actual data. 
Combining these reduction factor with the integrated luminosity for a month or
so, we obtain a
very conservative estimate for the expected number of events at RHIC
per year $ N_{events} \approx 10^6 $. We can expect, that an experimental
observable occuring with a likelihood larger than $10^{-6} $ can be
detected.

A simple method to estimate the production probability of charmed
hadrons is the coalescence model \cite{Dover90}. In its basic version
this model implies, that the production probability of a multi-quark
state should be proportional to the product of the ratios of the 
number densities of its
constituents normalized to the total density of the surrounding matter. 
In hot and dense hadronic matter charmed quarks are mainly 
formed via pair creation. The cross section for this pair creation
process for proton nucleus  collisions has been
measured \cite{FMNR94} and is in the order $\sigma_{c{\bar c}} \approx
0.04$ mb for a lab energy of $E_{lab} = 800$ GeV. Extrapolated to the
bombarding energy available at RHIC of $\sqrt{s} = 200$ GeV one gets
$\sigma_{c{\bar c}} \approx 300~\mu$b \cite{Vogt96}. This amounts to about 10 $c\bar c$ pairs
in one single central collision of gold nuclei at RHIC!
The charm pair production cross section has to be compared to the total
nucleon nucleon cross section of $\sigma_{NN} \approx 40$ mb. As a first
guess, we find that an exotic containing $C$ charmed quarks is
produced with a probability proportional to $ (\sigma_{c{\bar
c}}/\sigma_{NN})^C \approx 10^{-2C}$. 

This guess neglects the light quark content of the exotic. Actually,
in equilibrated hadronic matter the charmed quark--anti-quark pairs
would annihilate again and we would end up without any charmed
hadrons. Only the surrounding light quarks can combine with the
charmed quarks to form $D$--mesons, $\Sigma_c$--baryons etc. In this
way we maintain the charmed quarks, prevent their annihilation and
enhance the probability that these type of singly charmed objects
combine with itself or other hadrons to multi-charm and
multi-baryon number carrying exotics. High light quark density
is therefore essential to assure the production of these kind of exotics. 
The central rapidity region should be the
best area to find them experimentally.

An estimate of both, the light quark and baryon number density, 
is the number of nucleons $N_N$ of the collision
normalized by the number of pions $N_{\pi}$ expected for collisions at RHIC,
i.e.\ several thousand. The probability to find an charmed exotic 
with baryon number $A$ is therefore proportional to
the ratio $(N_N/N_{\pi})^A$.  At central rapidity this might be around
$10^{-2A}$. The suppression factor for 
forming a deuteron is then about $10^{-4}$. 

Putting these results together, we find that the production
probability of a charmed exotic with charm number $C$ and baryon
number $A$ is
\begin{equation}
P(C,A) = N_\pi \left(\frac{\sigma_{c{\bar c}}}{\sigma_{NN}}\right)^C
\left(\frac{N_N}{N_\pi}\right)^A
\approx 10^{3-2C-2A}
\label{eq:prob}
\end{equation}
From this estimate we gain a first feeling what possibly could be
detected at RHIC. Eq.~(\ref{eq:prob}) gives one deuteron in 10 events and 10
D-mesons per event. Combined with the supposed sensitivity of $10^{-6}$
the inequality 
\begin{equation}
  2 A + 2 C < 9
\end{equation}
for the detectability of multiply charmed exotics emerges.
A single charmed exotic carrying baryon number three,
a double charmed exotic with baryon number two, 
or a triple charmed exotic with baryon number one (the $\Omega_{ccc}$, the
charmed analog to $\Omega^-$),
seems to appear in 
enough events to be visible in the detectors.

\begin{figure}
\psfig{figure=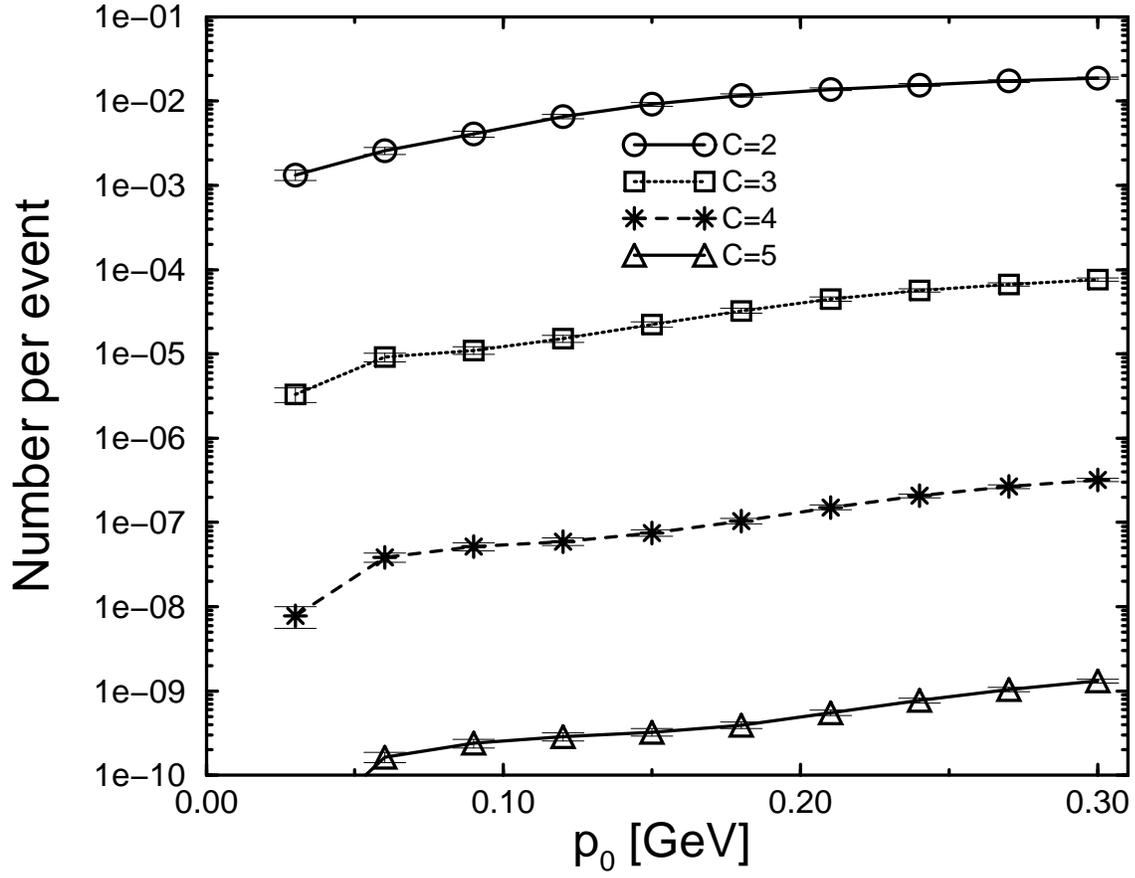,width=1.0\textwidth,angle=0}
\caption{Production probabilities for multiply charmed quark exotics at
RHIC ($s^{1/2}=200$ GeV) as a function of the coalescence parameter
$p_0$ with shadowing and a quench.}
\label{fig:prod1}
\end{figure}

\begin{figure}
\psfig{figure=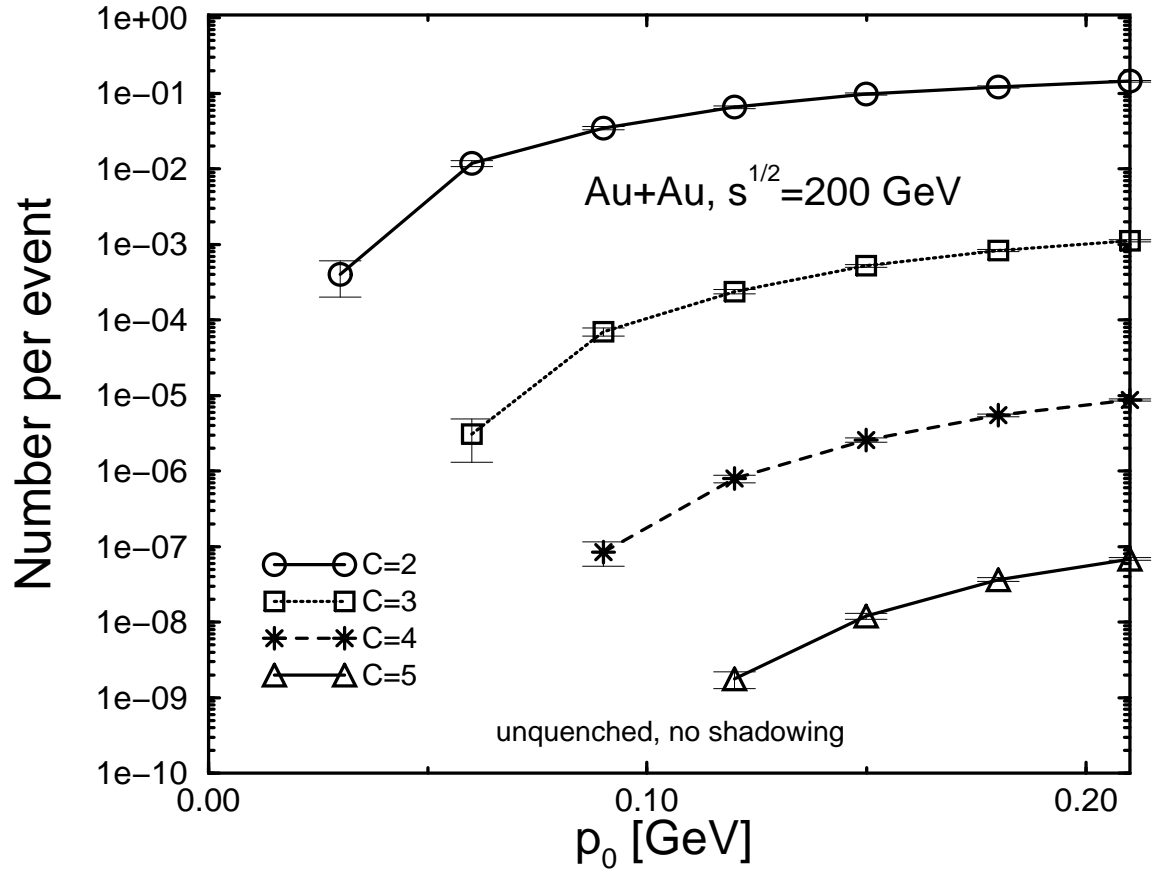,width=1.0\textwidth,angle=0}
\caption{Production probabilities for multiply charmed quark exotics at
RHIC as a function of the coalescence parameter
$p_0$ for the unquenched case with no shadowing.}
\label{fig:prod2}
\end{figure}

Let's solidify this result with a simulation. For a first estimate we used the
momentum distribution of charm quarks from the event generator HIJING
\cite{Lin97} as input for a simple coalescence model. Charm quarks sitting
in within a certain radius of momentum $p_0$ are supposed to form a multiply
charmed exotic. We assume, that the main penalty comes from the
coalescence of charm quarks and neglect the penalty of finding 
light quarks within the same area as they 
are abundantly produced at such high energy.
Recent investigations of charm quark production at RHIC indicate
a sizable energy loss of charm quarks during the expansion stage
especially in transverse momentum.
A so called quench incorporates the energy loss of the charm quarks by
rescattering with the surrounding particles after being produced in a heavy-ion
collision. 
This makes the phase space distribution of the charm quarks
narrower than in the unquenched case (for a more detailed 
discussion about this point see \cite{Lin97}). 
Fig.~\ref{fig:prod1} shows the number of produced charmed exotics for various
charm numbers as a function of the coalescence parameter $p_0$. 
Here we choose the case with shadowing and a quench, 
the case for no shadowing without a quench is shown in Fig.~\ref{fig:prod2}.
In both cases, the production of double and triple charmed objects seems to be
detectable at RHIC, i.e.\ the production rate is above $10^{-6}$ per event.
A quench enhances the formation of charmed matter as the heavy
quarks are closer together in phase space.
We find indeed, that for the quenched case, 
the production rate is higher and less sensitive to the coalescence
parameter $p_0$ as the charm quarks.
Another scenario was recently discussed in \cite{SU97} which will also enhance
the production rates of charmed matter droplets.
Within a cascade hydrodynamical model, 
they find that charmed particles feel drag and diffusion forces within
quark-gluon plasma droplets so that the charm quarks 
resides inside the droplet.
This mechanism resembles very much the separation mechanism for strange
quarks: the quark-gluon plasma droplet gets enriched with strangeness
as it is energetically more favorable to keep the strange quark inside the
droplet \cite{Greiner91}. 
Here we point out that the distillation mechanism proposed for the formation of
strangelets \cite{Greiner88} will
also work for charmed matter, as the D-meson is much lighter than the lightest
baryon with charm $\Lambda_c$. This effect will furthermore enhance the charm
fraction of the droplet during cooling down by emitting D-mesons and carrying
away anticharm quarks, therefore enriching the plasma droplet with charm.
Nevertheless, this implies a baryon-rich regime
or strong fluctuations of baryon number which one might expect at RHIC energy
\cite{Spieles96}. 
Therefore, 
charmed exotics of $|C|>3$ might well be produced in measurable amounts
if a quark-gluon plasma forms. Hence, we will investigate in the following the
properties of charmed objects for $|C|>3$, too.


\section{Charm Matter}

First, let us discuss the properties of charm matter in bulk.
Strange matter gets enhanced stability compared to two flavor matter 
by introducing a new degree of freedom and a lowering of the Fermi energy
\cite{Chin79}. 
This is possible, as the strange quark mass is still light enough.
For the charm quark with a mass of the order of 1.5 GeV, the Fermi energy can
not be lowered at reasonable density. 
Calculations for hybrid stars with charm quarks conclude that this quark is
probably too
heavy to appear in the core of a neutron star \cite{Kettner95}.
On the other hand, the potential generated by an effective
one-gluon exchange gets attractive and gives an enhanced stability for 
heavy quarks. 
This potential is repulsive for massless quarks and is evident from
the energy density for massless quarks which reads
\begin{equation}
\epsilon_0 = \frac{3}{2\pi^2} \mu_0^4 \left(1+ \frac{8\alpha_c}{\pi}\right) 
\end{equation}
Here $\alpha_c$ is the strong coupling constant and $\mu_0$ is the chemical
potential of the up and down quarks.
Note that for values of $\alpha_c>\pi/8\approx0.39$ the correction from
one-gluon exchange is bigger than the zeroth order. 
Usually, bag model fits to hadron spectra yield $\alpha_c>\pi/8$
\cite{DJJK75,Izatt82} indicating a highly non-perturbative potential.
This is a problem for bulk matter, as the pressure gets negative
because
\begin{equation}
  p_0 = \frac{\mu_0^4}{2\pi^2}\left(1-\frac{8\alpha_c}{3}\right)
\end{equation} Therefore, the parameters extracted from a fit to hadron spectra
can not be used for bulk matter calculation. This indicates that the bag model
is an effective model and that the parameters change when going to higher
density and/or strangeness numbers. 

\begin{figure}
\psfig{figure=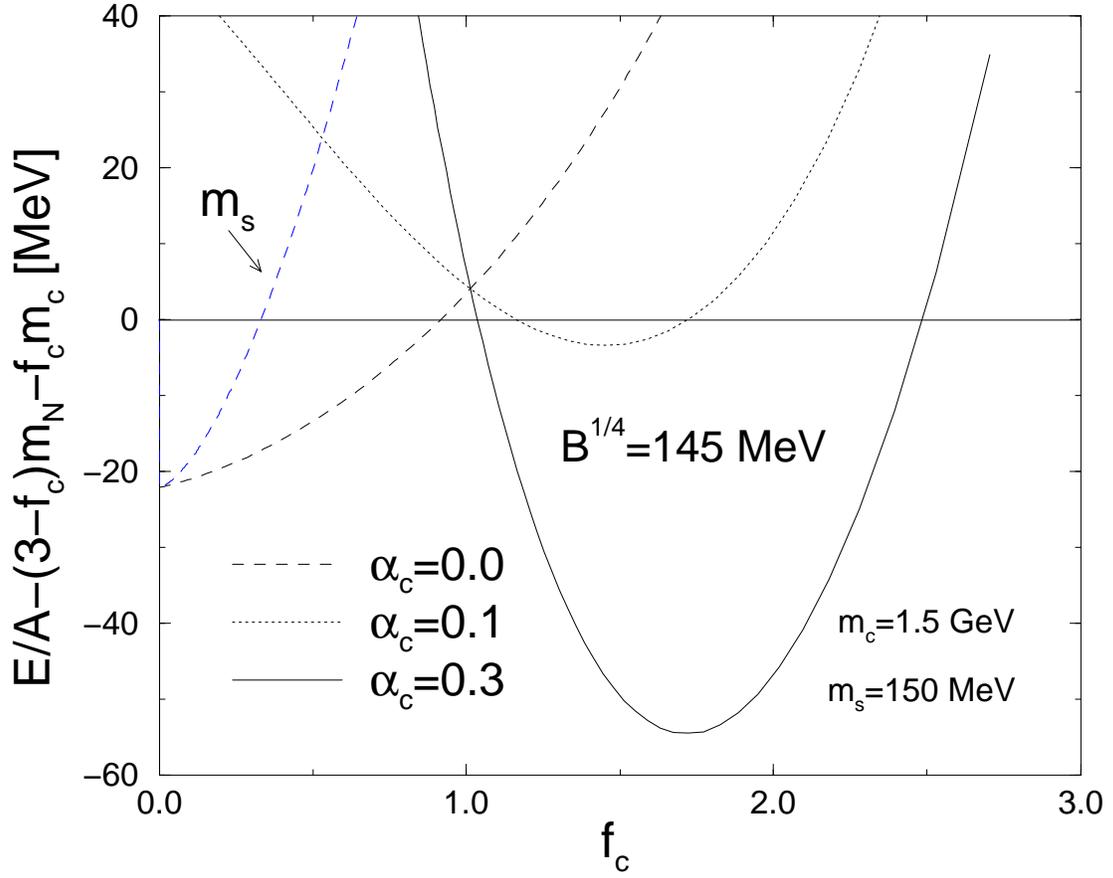,width=1.0\textwidth,angle=0}
\caption{Binding energy of charmed matter for various strong coupling
constants. For comparison the case for strange matter indicated by $m_s$ is
also shown.}
\label{fig:energy}
\end{figure}

The situation changes for a finite quark mass.
The energy density for quarks with a finite mass is given by \cite{Free78}
\begin{equation}
\epsilon_i = \frac{3}{8\pi^2} \left[ \mu_i^2 \left(2\mu_i^2-m_i^2\right) -
m^4_i \log \frac{\mu_i+x}{m_i}
+ \frac{16\alpha_c}{3\pi}
\left\{ 3\left(\mu_i x- m_i^2 \log \frac{\mu_i+x}{m_i}\right)^2
-2 \left(\mu_i^2-m_i^2\right)^2 \right\} \right]
\end{equation}
with $x=\sqrt{\mu_i^2-m_i^2}$.
Here the interaction gets attractive for large enough quark mass $m_i$.
This can be seen in Fig.~\ref{fig:energy}, where the total energy per baryon
number (minus the mass) is plotted versus the charm fraction $f_c=|C|/A$
for increasing values of the strong coupling constant $\alpha_c$. 
For $\alpha_c=0.3$, charm matter is bound by 55 MeV at $f_c=1.7$. 
Hence, the charm quark is heavy enough to feel an attractive interaction from
the effective one-gluon exchange. 
For comparison, the case for strange matter indicated by $m_s$ is also shown,
then as a function of the strangeness fraction $f_s=|S|/A$.
Only the case $\alpha_c=0$ can be seen on the plot. Adding the one-gluon
exchange contribution results in a repulsive term, so that the curves for the
strange quarks are shifted up and are lying outside the range displayed in the
figure. 

\begin{figure}
\psfig{figure=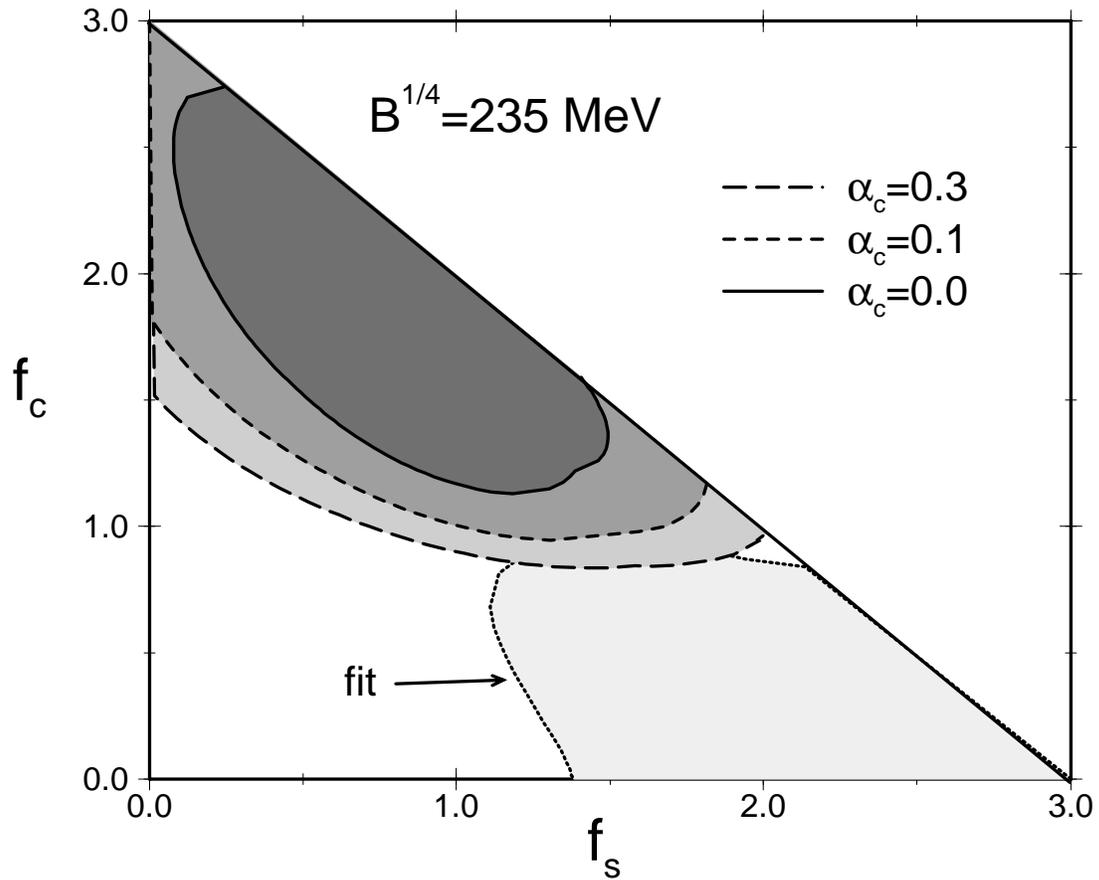,width=1.0\textwidth,angle=0}
\caption{Allowed region for the formation of charmed matter as a
function of strange and charmed quark fraction, 
$f_s$ and $f_c$ respectively. The
different shaded regions correspond to different values of the
strong coupling constant $\alpha_c$. The region labeled 'fit'
corresponds to our bag fit of $\alpha_c=0.476$ and $B^{1/4}=168$ MeV.}
\label{fig:matter}
\end{figure}

In general, charm matter consists of strange quarks also. We plot the areas,
where charm matter is bound in a contour plot as a function of the charm and
strangeness fraction in Fig.~\ref{fig:matter} for different values of
the strong coupling constant $\alpha_c$. Note that charm matter can not exist
for $f_s+f_c>3$ by definition (one can not have more than three quarks per
baryon number). The calculation was done for a bag parameter of $B^{1/4} = 235$
MeV taken from a fit to charmonium levels \cite{Hasen80}
which is in accordance with estimates from QCD sum rules \cite{SVZ79}.
Also shown is the case for the parameters taken from a fit to the hadron
spectra (denoted by fit) which we will discuss in the following section.
This is the area in the lower right corner of the figure with a much lower bag
parameter of $B^{1/4}=168$ MeV. 
Note that the
pressure is negative for the massless quarks in this case as
$\alpha_c=0.476>\pi/8$! 

Charm matter is bound when the total energy per baryon is lower than 
the sum of the corresponding hadron masses. The masses of the hadrons with
$C\geq 2$ are not known and taken from the bag model
calculation (see Table~\ref{tab:masscharm}). 
Strange matter along the line for $f_c=0$ 
is unstable for this bag parameter \cite{Greiner88} except for the parameters
taken from the fit. 
Nevertheless, charm matter demonstrates to be stable for a wide range of 
charm and strangeness fraction of $f_c=1-3$ and $f_s=0-1.5$ even for this high
bag parameter. Hence, charm matter might be stable while pure strange matter is
not. The next step is to calculate the masses of finite charm matter for low
baryon numbers. This will be done in the following sections after introducing
the modified bag model for multiply charm quark bags.


\section{Modified Bag--Model}

At present our technical skills in solving QCD do not suffice to
accurately predict boundstate energies of multi--quark states. Instead
we have to rely on simple model calculations and estimates. A
very simple, controllable and relatively reliable model for our purposes is
the bag model \cite{DJJK75}. We think the results of this model give
us a qualitative first impression of the stability of multi--quark
states. More detailed calculations can be performed both through
potential interactions and on the lattice.

We modify the commonly used version of the bag model in
two essential ways. First, we evaluate the bag energy by setting the
strong coupling $\alpha_c$ to zero \cite{Izatt82,Zouzou94}, effectively
neglecting all color electric and magnetic contributions. Both
contributions are evaluated after minimization of the
bag energy, then added to the bag mass, and therefore treated as 
true lowest order corrections. 
In this way we
take only the gluon exchange contribution into account and our bag is
locally color charged, while globally color neutral. We postulate that
all other color electric and color magnetic 
corrections can be accounted for by
adjusting the parameters of the model. 
And secondly,
we treat all quarks on the same footing, i.e.\ we do not distinguish
between heavy and light quarks, rather demand flavor symmetry which is
broken by the explicit masses of the quark species.
This simplifies our approach. A very
heavy quark, like the bottom is thus not treated non--relativistically
and is not sharply localized at the center of the bag. Instead it is
distributed, with the maximum of the distribution being closer to 
the center of the bag than the one of the light
quarks. Both of these modifications are ad hoc and can only be justified by
the success of the phenomenological model in describing the
experimentally measured hadronic masses.

We are interested in a first crude estimate of the binding energy of a
multi--quark state with $n$ quarks. All quarks are confined inside a
spherical bag of radius $R$. The energy or mass of such a state is to zeroth
order in the strong coupling $\alpha_c$ given by
\begin{equation}
E_n = \sum_{i=1}^n\,\omega_i -\frac{Z_0}{R} +
\frac{4}{3}\,\pi\,B\,R^3\,,
\label{bagE}
\end{equation}
where $\omega_i$ is the eigenenergy of the $i$-th quark
confined within the bag, $Z_0$ is the Casimir energy of the
bag \cite{Don80} absorbing all effects $\sim 1/R$ 
and $B$ is the bag constant. Minimization of the bag
energy determines the bag radius $R$.

\begin{figure}
\psfig{figure=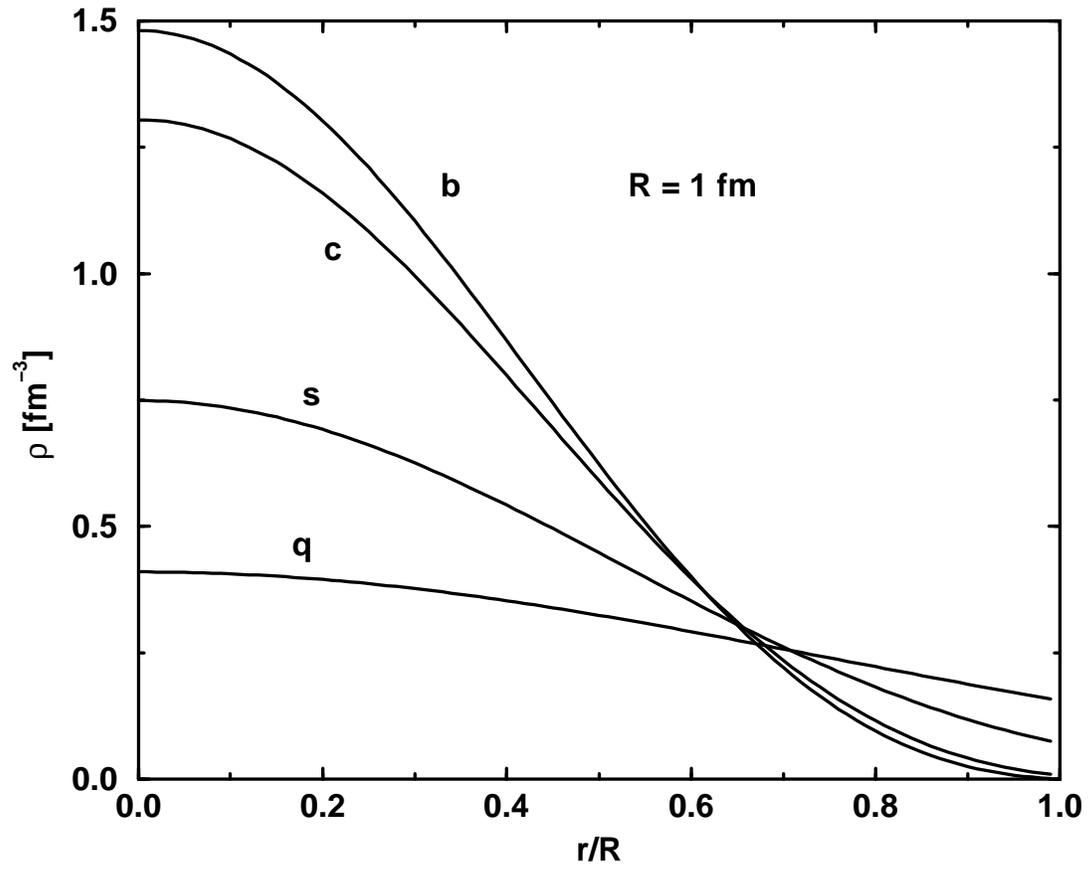,width=1.0\textwidth}
\caption{Probability density of a quark of flavor $q(u,d),s,c,b$ for a
bag of radius $R=1$ fm as a function of the scaled radius.}
\label{fig:density}
\end{figure}

The confined $i$-th quark has to satisfy both the free Dirac equation
and the linear boundary condition at the bag radius $R$
which leads to the constraint for the single particle energy $x_i$ of a quark 
\begin{equation}
\tan x_i = \frac{x_i}{1-R\,(\omega_i+m_i)}\,,
\label{x}
\end{equation}
where $m_i$ is the mass and $\omega_i$ the total energy.
In Fig.~\ref{fig:density} we plot the probability density $\rho$
of a quark of flavor
$q(u,d),s,c,b$ for a
bag of radius $R=1$ fm as a function of the scaled radius $r/R$. The
probability density is proportional to the sum of the square of 
the spherical Bessel functions
\begin{eqnarray}
\rho ({\bf r}) &=& \Psi_i({\bf r},t)^{\dagger}\gamma_0\Psi_i({\bf r},t)
\nonumber \\
&=& \frac{N^2}{4\pi\omega} \left( (\omega+m) j_0(x_i\,r/R)^2
+(\omega-m) j_1(x_i\,r/R)^2 \right)\,.
\end{eqnarray}
The heavier the quark, the more localized is the density at $r=0$ and
the smaller are the values for the spherical Bessel Functions.

The minimization of the bag energy in eq.~(\ref{bagE}) fixes the bag
radius $R$. We use this radius to determine the color electric and color
magnetic corrections to order $\alpha_c$. Both corrections were
discussed by many authors, e.g.\ \cite{DJJK75,Izatt82,Zouzou94} 
and we quote only the results. The color electric correction $V_E$ is
\begin{eqnarray}
V_E &=& \sum_{i>j}\sum_a(\lambda^a_i\cdot\lambda^a_j)\times E_{ij}
\nonumber\\
&=&\sum_{i>j}\sum_a(\lambda^a_i\cdot
\lambda^a_j)\,
\frac{\alpha_c}{2\,R}\,\int_0^1\,\frac{du}{u^2}\,\rho_i(u)\,\rho_j(u)\,,
\label{delel}
\end{eqnarray}
where the $\lambda^a_i$ are the color matrices of particle $i$ 
and the density $\rho_i$
is given by
\begin{eqnarray}
\rho_i(u) = \frac{\omega_i\,[x\,u-\sin^2(x\,u)/x\,u]-
m_i\,[\sin(x\,u)\,\cos(x\,u)-\sin^2(x\,u)/x\,u]}{\omega_i\,[x-\sin^2(x)/x]
-m_i\,[\sin(x)\,\cos(x)-\sin^2(x)/x]}\,.
\end{eqnarray}
Note, that
the summation is performed over all quark pairs in the bag ($i>j$). 
In the original MIT bag model, the sum runs over all quark combinations ($i,j$)
\cite{DJJK75}. 
Then, the color electric potential for the charmonium state will vanish
\begin{equation}
V^{old}_{c\bar c} 
= \sum^2_{i,j}\sum_a(\lambda^a_i\cdot\lambda^a_j)\times E_{cc}
= C_3[1] \times E_{cc} = 0 
\end{equation}
as it is a color singlet state for which the Casimir operator of color SU(3) 
$C_3[1] = 0$. Therefore, the original MIT
bag model can not describe the mass of J/$\Psi$, as it is mainly bound by the
Coulomb-like color electric potential. In our case, the sum runs only over
unequal quark pairs ($i>j$), therefore, the color electric potential turns out
to be 
\begin{equation}
V_{c\bar c} = \sum^2_{i>j}\sum_a(\lambda^a_i\cdot\lambda^a_j)\times E_{cc}
= \left(\frac{1}{2}C_3[1]-\frac{16}{3}\right) \times E_{cc} = -\frac{16}{3}
E_{cc}  
\end{equation}
i.e.\ nonvanishing and strongly attractive, as needed to describe the masses of
$J/\Psi$ and $\eta_c$ correctly. 
This procedure allows now 
for a {\em local} color charge (there is an intrinsic color charge inside the
bag),  but the overall color charge still
vanishes as the sum over all quarks in the bag gives
\begin{equation}
\sum_k\sum^{N_k}_{i,j}\sum_a(\lambda^a_i\cdot\lambda^a_j) = 
\sum_k C^k_3 = C^{total}_3 = 0 \quad .
\end{equation}
Here $C_3^k$ is the Casimir operator of the quark flavor $k=q(u,d),s,c$.
Hence, light quarks and heavy quarks can have a color charge and do not need to
be color neutral separately, but of course the sum of all color charges gives
an overall color neutral object. This new color degree of freedom will be
important for the enhanced stability of multiquark states which we will
discuss later in more detail. 

The color magnetic correction $V_M$ is
\begin{eqnarray}
V_M &=&-\sum_{i>j}\sum_a({\bf \sigma}_i\lambda^a_i)\cdot({\bf 
\sigma}_j\lambda^a_j)\,M_{ij}
\nonumber\\
&=&-\sum_{i>j}\sum_a ({\bf \sigma}_i\lambda^a_i)\cdot({\bf 
\sigma}_j\lambda^a_j)\,
\frac{3\,\alpha_c}{R}\,\frac{\mu(m_i,R)}{R}\,\frac{\mu(m_j,R)}{R}\,
\left(1+2\,\int_0^R\,\frac{dr}{r^3}\,\mu(m_i,r)\,\mu(m_j,r)\right)\,,
\label{delmag}
\end{eqnarray}
where the $\sigma_i$ are the spin matrices and the magnetic moment 
$\mu_i$ of the quark eigenmode given by
\begin{eqnarray}
\mu_i(m_i,r) = \frac{1}{6}\,\frac{4\,r\,\omega_i+2\,r\,m_i-3}{2\,
\omega_i\,(r\,\omega_i-1)+m_i}\,.
\end{eqnarray}
The summation runs again over all quark pairs in the bag.

In Fig.~\ref{fig:electric} 
we plot the electric interaction $E_{ij}$ 
and the magnetic interaction $M_{ij}$
between two quarks with flavor $i,j=q(u,d),s,c$
as a function of the bag radius $R$. We see that the largest
contribution to the total electric
interaction energy within the bag is given by the interaction between two
heavy quarks. In contrast, the largest
contribution to the total magnetic
interaction energy within the bag is given by the interaction between two
light quarks. A multi-quark bag becomes therefore more stable the more
heavy charmed quarks we add, due to the increase in the electric interaction
energy, and it becomes more stable the more flavor antisymmetric we
arrange the light quarks (u,d,s). This property will allow us to
simplify the study of the group structure of multi--quark bags in the
next section.

As a test of our modified bag model we calculate first the measured
masses of the light hadrons built up from $u,d$ and $s$ quarks. First,
we use some of the known masses to fix the parameters of
the model, 5 in the case of the light hadrons. The
mass of the lightest quarks is set to zero $m_q=0$, where $q=u,d$. The
strong coupling constant $\alpha_c = 0.4764$ is fixed from the 
color magnetic mass
splitting of the proton $p$ and the delta $\Delta$. These two
particles have identical color electric interactions . The masses of
the proton and the mass of the degenerate $\rho, \omega$--system 
are used to fix
the bag constant $B^{1/4} = 0.1684$ GeV and $Z_0 = 1.585$. 
Finally, we use the $\Omega^-$ hyperon to fit the strange quark
mass $m_s = 0.342$ GeV. The various color and spin assignments of the
color electric and magnetic energies in (\ref{delel}) and
(\ref{delmag}) are given in \cite{DJJK75}. 

In Table~\ref{tab:masses} we list our results for the light
hadrons. We reach satisfactory agreement for the heavier hadrons, but
have problems with the lightest mesons like the $\pi,\eta$ system and
light strange mesons like the $K$ and the $K^{\star}$. We attribute
this to their role as Goldstone bosons of chiral symmetry. The
slightly larger error of the $\phi$ might be due to its non--spherical shape.

In the next step we include the two heavier quarks and check
if our model continues to perform
reasonably well. The charm mass $m_c = 1.788$ GeV and the
bottom mass $m_b = 5.296$ GeV are fitted to the mass of the
$\Lambda_c$ and the $B_s$ respectively. With these quark masses fixed
we can predict all other hadrons containing $u,d,s,c$ or $b$
quarks. We find in Tables~\ref{tab:masscharm} and \ref{tab:massbeauty}
good agreement between the experimental measured
masses and our masses. In most cases we have a relative deviation from
the measured masses of less than $2-3\%$. 
The predicted 
masses of $\Omega_{cc}$ and $\Omega_{ccc}$ are a little bit higher than 
estimates from potential models \cite{Martin95}.
With this in mind we can
proceed in the next section to analyze the color and spin assignments
for multi-charmed exotics and use then our modified bag model to
predict their masses.


\section{Possible Candidates and their masses}

\subsection{Estimate of the color magnetic potential}

\begin{figure}
\psfig{figure=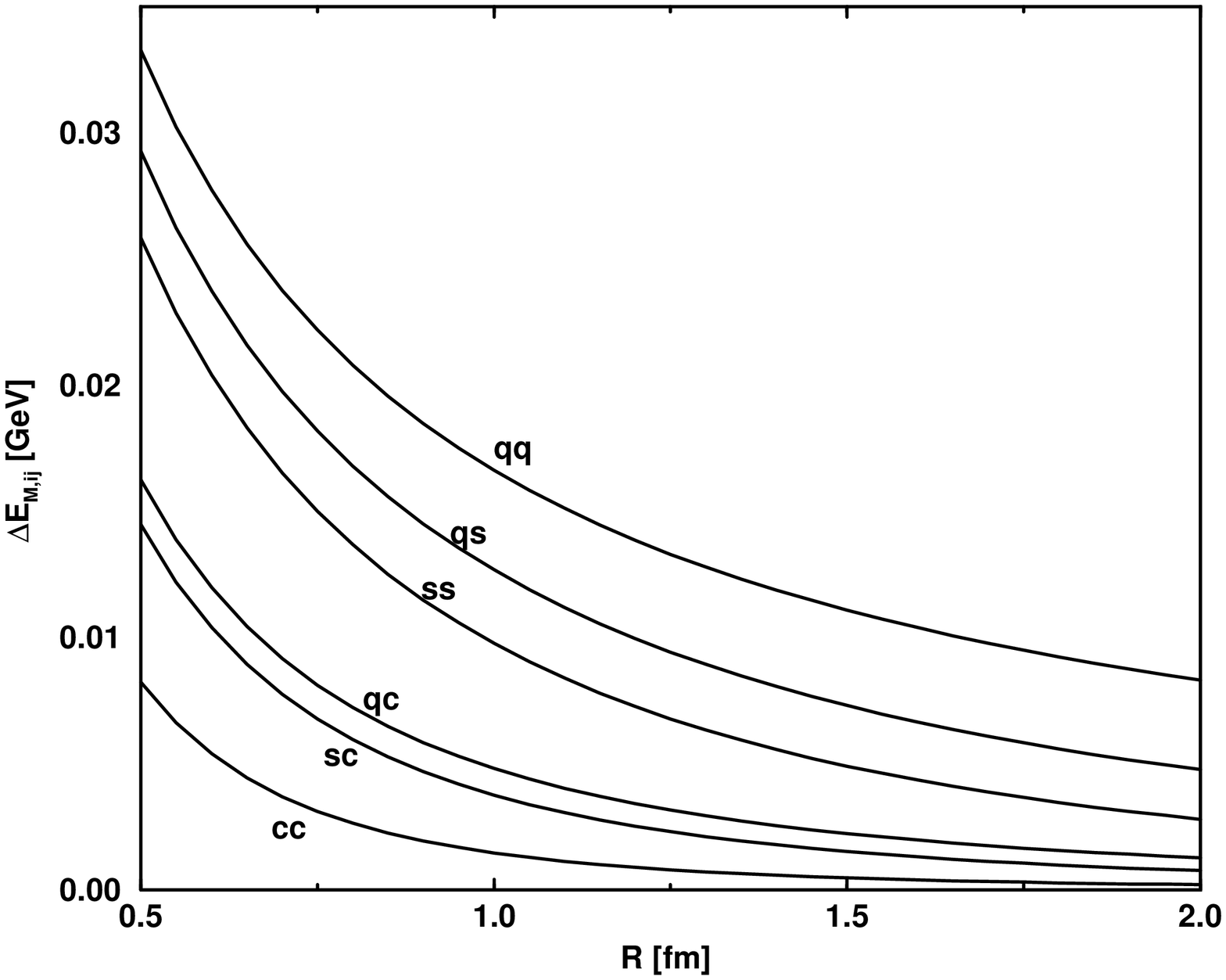,width=1.0\textwidth,angle=-90}
\caption{Magnetic interaction $M_{ij}$
between two quarks with flavor $i,j=q(u,d),s,c$
as a function of the bag radius $R$.}
\label{fig:magnetic}
\end{figure}

As seen in the previous section, quark bags can be bound by either the color
magnetic force or the color electric force. 
The former one is dominant for light quarks, while the latter one is most
pronounced for the heavy quarks c and b (see Figs.~\ref{fig:electric} and
\ref{fig:magnetic}). For charmlets, one can show that both forces are
attractive! Let us start with the color magnetic term.
Here we follow the discussion outlined by Jaffe \cite{Jaffe91} for strangelets
but with an unprecedented conclusion about charmlets.
In this subsection, 
we consider up, down and strange quarks as light quarks and discuss effects due
to the finite strange quark mass later. 

Three light quarks 
can be combined in SU(6)
color-spin to a [56]-plet, two [70]-plets and a [20]-plet. A decomposition into
$SU(3)_{color}\times SU(2)_{spin}$ shows that there is a color singlet for the
[70]-plet (the baryon octet) and for the [20]-plet (the baryon decuplet), but
not for the [56]-plet. The color magnetic term is proportional to
\begin{equation}
V_M(d_6,d_3,S,N_q) = 
\sum_{i>j}\sum_a (\sigma_i\cdot\lambda^a_i)(\sigma_j\cdot\lambda^a_j) 
\; M_{ij}  =
\left( 8N_q - \frac{1}{2} C_6[\mu] + \frac{1}{2} 
C_3 + \frac{4}{3} S(S+1) \right) \; M_{ij}
\label{eq:lalamag}
\end{equation}
where $N_q$ is the number of light quarks, 
$C_6$ is the Casimir operator of SU(6) which
depends on the representation $[\mu]$, $C_3$ is the Casimir operator of
$SU(3)_{color}$, and $S$ denotes the total spin. $M_{ij}$ is the matrix
element defining the strength of the interactions. The potential depends on the
dimensions of color-spin SU(6) and color SU(3) following the notation of 
\cite{HS78}. 
From eq.~(\ref{eq:lalamag}) one sees, that the representation with the highest
Casimir operator $C_6$ will be mostly bound. Generally, the representation
which is mostly symmetric in color-spin will have this feature
\cite{Jaffe77}. It turns out that the [56]-plet is the most symmetric
and will be more bound than baryons. But this one is color charged and does not
exist in nature then. Now we include heavy quarks, like the charm quark, which
will have a negligible contribution to the color magnetic term due to its heavy
mass. Then one can make the [56]-plet of light quarks color neutral by adding
e.g\ three charm quarks (or antiquarks) and gain this additional binding
energy. This is of course in line with the discussion about the hyperfine
interaction for pentaquark states introduced by Lipkin \cite{Lipkin87} 
but from a slightly 
different point of view. One can estimate the binding energy by assuming that
the matrix element $M_{i,j}$ is the same for all light quarks (i.e.\ setting
$m_s =m_{u,d}$) and independent of the radius of the bag. 
Then one gets for the binding energy
\begin{eqnarray*}
B(N_q=3) &=& V_M(56,8,1/2,3) - V_M(70,1,1/2,3) 
= -14M + 8M \cr
&=& -6M = -6\times\frac{(M_\Delta - M_N)}{16}
\approx -110 \mbox{ MeV}
\end{eqnarray*}
The necessary values for the 
Casimir operators can be found in \cite{HS78}.
We note that the [56]-plet is a singlet in flavor, as overall Fermi statistics
require that the most symmetric one in SU(6) color-spin must be most
antisymmetric in SU(3) flavor.
Hence, the proposed exotic
states would have a quark content of \{udsccc\} (the $H_{ccc}$ dibaryon) 
and \{uds\=c\=c\=c\} (similar to a bound state of two D-mesons and one D$_s$
meson). 

The same line of argumentation can be repeated for four light quarks, which are
in a [210]-plet and are color neutral by adding an anticharm quark (this is the
well known pentaquark $P_{cs}$ \cite{Lipkin87}) 
or two charm quarks (the $H_{cc}$ dibaryon with \{uudscc\}, \{uddscc\}) or
=\{udsscc\}). The binding energy is then
\begin{eqnarray*}
B(N_q=4) &=& V_M (210,3,0,4) - V_M(70,1,1/2,3)
= -16M + 8M \cr
&=& -8M =  -\frac{1}{2} \left(M_\Delta - M_N\right)
\approx -150 \mbox{ MeV}
\end{eqnarray*}
The $H_{cc}$ has been also discussed in \cite{Chow95}.
For five light quarks, one gets
\[
B(N_q=5) = V_M (420,3,1/2,5) - V_M(70,1,1/2,3) - V_M(21,3,0,2)
= -16M + 16M = 0
\]
using $C_6[420]=358/3$ which we computed from the Tables in \cite{IN66} and
the formula given by Jaffe \cite{Jaffe77b}.
This does not support the possibility 
for the bound exotics $H_c$. For the heptaquarks
\{uudds\=c\=c\}, \{uudss\=c\=c\}, \{uddss\=c\=c\} the situation  
is different as
\[
B'(N_q=5) = V_M (420,3,1/2,5) - V_M(70,1,1/2,3)
= -16M + 8M = -8M \approx -150 \mbox{ MeV} 
\]
as the two charm antiquarks combine to two D-mesons without any gain in the
color magnetic energy.
The case for $N_q=6$ gives
\begin{eqnarray*}
B(N_q=6) &=& V_M (490,1,0,6) - 2\times V_M(70,1,1/2,3) 
= -24M + 16M = -8 M \approx -150 \mbox{ MeV}
\end{eqnarray*}
The line of argumentation stops here as it is just the famous case for the
H-dibaryon \cite{Jaffe77} which is color neutral by itself. Hence, the exotics
\{uuddssccc\} and \{uuddss\=c\=c\=c\} only exist, if the H-dibaryon
is bound and/or the interaction between heavy and light quarks is attractive
enough.

\subsection{Estimate of the color electric potential}

\begin{figure}
\psfig{figure=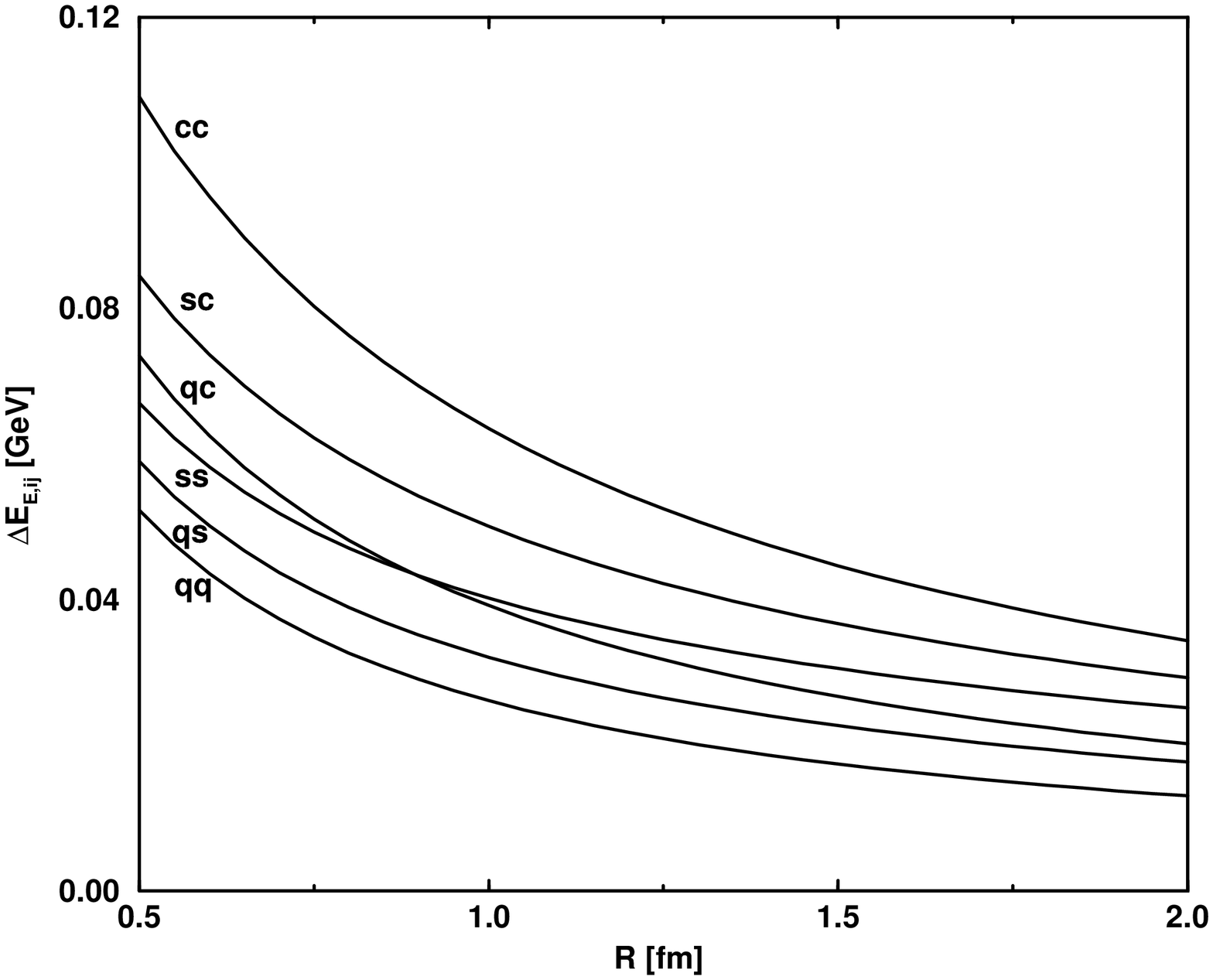,width=1.0\textwidth,angle=-90}
\caption{Electric interaction $E_{ij}$
between two quarks with flavor $i,j=q(u,d),s,c$
as a function of the bag radius $R$.}
\label{fig:electric}
\end{figure}

The discussion of the color electric term 
is simpler, as it involves only color SU(3). Between the charm quarks one finds
the color electric potential
\begin{equation}
V_E(d_3,N_c) = \sum_{i>j}\sum_a (\lambda^a_i \cdot \lambda^a_j) 
\; E_{ij} = \left(\frac{1}{2} C_3^c - \frac{8}{3}N_c \right) E_{cc}
\label{eq:lalael}
\end{equation}
where $E_{cc}$ is the matrix element for the color electric interaction.
This is the Coulomb-like potential used in e.g.\ charmonium potential 
models. This potential increases linearly with the number of heavy quarks hence
giving an enhanced stability for multiply charmed objects!
In addition, we calculate also the color electric potential between the charm
quark and the light quarks:
\begin{equation}
V'_E(d_3,N_q,N_c) = \left( 
\sum^{N_q+N_c}_{i>j} \sum_a (\lambda^a_i \cdot \lambda^a_j) 
-\sum^{N_q}_{i>j} \sum_a (\lambda^a_i \cdot \lambda^a_j) 
-\sum^{N_c}_{i>j} \sum_a (\lambda^a_i \cdot \lambda^a_j) 
\right) E_{ij} 
= - C^c_3 E_{qc}
\label{eq:lalael2}
\end{equation}
as the contribution is quite sizable (see Fig.~\ref{fig:electric}).
Interestingly, this interaction does not depend on the number of light and
heavy quarks. Its strength is determined by the internal color structure and
it is always attractive.
Note that the quark bag is still {\em globally} color neutral but {\em locally}
color charged. 

The matrix element increases with increasing mass as evident in
Fig.~\ref{fig:electric} so that the tetraquark with two heavy quarks $Q$ and
two light antiquarks is bound in the limit of $m_Q\to\infty$.   
The case for the tetraquark with charm (cc\=q\=q) is still under debate
but the bottom quark seems to be heavy enough to support a bound
state \cite{Carlson88}.

\subsection{Candidates for the flavor SU(3) symmetric case}

With the preceding discussions about the main color magnetic and electric
potential terms, one can now look for possible candidates in the flavor SU(3)
symmetric case, ignoring a finite mass of the strange quark mass.

We calculate the coefficients in front of the matrix elements for the magnetic
interaction between the light quarks, for the electric interaction between the
heavy quarks and the one between heavy and light quarks.
Bags up to 6 light quarks and 6 heavy quarks are considered. The result is
shown in Table~\ref{tab:cand} for charm quarks and in Table~\ref{tab:anticand}
for the case of anticharm quarks. The matrix elements are compared to the
possible decay products on the right side of the tables. 
We consider also the two possible color representations. The color-spin
multiplets can be decomposed into $SU(3)_{color}\times SU(2)_{spin}$ as
\begin{eqnarray*}
N_q=2:&\qquad &[21] \supset (3,0) + (\bar 6,1) \cr
N_q=4:&\qquad &[210] \supset (3,0) + (\bar 6,1) + \dots \cr
N_q=5:&\qquad &[420] \supset (\bar 3,1/2) + (\bar 6,1/2) + \dots \cr
N_q=6:&\qquad &[490] \supset (1,0) + (8,1) + \dots 
\end{eqnarray*}
For $N_q=3$, 
we take the [56]-plet for the color octet and the [70]-plet for the
color singlet representation, respectively.

One notes, that for certain candidates, like \{3q3c\}, 
the coefficients for the
potentials for the candidate and its decay product are just the same.
Hence, these candidates are not likely to be bound. 
In four cases, only 
the color magnetic term is larger for the candidate than for the
decay products (\{cc4q\},\{6c6q\}, the pentaquark \{\=c4q\}, and \{6\=c6q\}). 
Other candidates get additional binding energy from the magnetic term and one
of the color electric terms depending on the color charge representation.
For the cases of different color states for the charm quarks, one observes that
the one with the higher dimension in color SU(3) gives less binding for the
magnetic term and the electric term $E_{cc}$ but more binding for the electric
term between the heavy and the light quarks $E_{qc}$. 
As the latter term is the most dominant
contribution to the overall binding energy (see Tables~\ref{tab:cand} and
\ref{tab:anticand}), it might well be that multiply
charmed objects are bound in a higher representation of color SU(3)!

To clarify this, we estimate the matrix element from
Figs.~\ref{fig:electric} and \ref{fig:magnetic} to $M=18$ MeV, $E_{qc}=50$ MeV
and 
$E_{cc} = 70$ MeV and calculate the binding energy. Table~\ref{tab:esti} shows
the result. Indeed, for some cases the higher color charge representation turns
out to be more stable: 
for \{3c6q\}, \{\=c\=c2q\}, \{\=c\=c5q\}, and \{\=c\=c\c6q\}.
Accidentally, it comes out to be the same for two cases (\{cc4q\} and
\{4\=c4q\}). Note, that we predict on this basis, that the tetraquark  
\{\=c\=c\c2q\} will be more bound, if the charm quark is sitting in the higher
color SU(3) representation, i.e.\ in the color 
sexplet instead of the usually assumed color triplet. The tetraquark is then
bound by the electric interaction between the heavy and light quarks and not
by the Coulomb-like potential between the heavy quarks!
The binding energy is quite sizable but will be diminished by flavor SU(3)
breaking effects and finite size effects due to the different radii of the
candidate and the decay products.

\subsection{Flavor breaking and finite size effects}

For the flavor breaking case, one has to calculate the
coefficient for every quark-quark combination.  
In the following, we take into account the finite strange quark mass and
consider three different quarks species: $q(u,d),s,c$.
 
The color electric contribution between the same quark species in the color
state defined by $C^k_3$ is given by
\begin{equation}
e_{kk} = 
\sum^{N_k}_{i>j}\sum_a (\lambda^a_i \cdot \lambda^a_j) =
\frac{1}{2} C_3^k - \frac{8}{3} N_k
\end{equation}
where the sum runs only over the quark species $k$.
Hence, for the charmonium one gets $e_{cc} = -16/3$, but for a baryon with two
charm quarks it is $e_{cc} = C_3[3]/2 - 16/3 = -8/3$, a factor of two lower,
often encountered in potential models.
The case of three different quark species can be found by using
the definition of the quadratic Casimir operator for color SU(3)
\begin{equation}
\sum^{N_k+N_l}_{i,j} \sum_a (\lambda^a_i \cdot \lambda^a_j) =
\sum^{N_k}_{i,j} \sum_a (\lambda^a_i \cdot \lambda^a_j) +
\sum^{N_l}_{i,j} \sum_a (\lambda^a_i \cdot \lambda^a_j) +
2\sum^{N_k+N_l}_{i>j} \sum_a (\lambda^a_i \cdot \lambda^a_j) =
C_3^{k+l} = C_3^m
\end{equation}
where the Casimir operator of color SU(3) for the two quark species is just the
one of the remaining quark species $m$ due to color neutrality.
This results in
\begin{equation}
e_{kl} = 
\sum^{N_k+N_l}_{i>j} \sum_a (\lambda^a_i \cdot \lambda^a_j) =
\frac{1}{2} \left(
C_3^m - C^k_3 - C^l_3 \right)
\end{equation}
The same procedure can be done for the color magnetic interaction, here for 
SU(6). Defining
\begin{equation}
a_{kl} = 
\sum^{N_k+N_l}_{i>j}\sum_a 
(\sigma_i\cdot\lambda^a_i)(\sigma_j\cdot\lambda^a_j) 
\end{equation}
gives
\begin{equation}
a_{kk} = 8N_k - \frac{1}{2} C^k_6[\mu] + \frac{1}{2} 
C^k_3 + \frac{4}{3} S(S+1) 
\end{equation}
for equal quarks, otherwise
\begin{equation}
a_{kl} = -\frac{1}{2} \left( C_6^{k+l} - C_6^k - C_6^l \right)
+ \frac{1}{2} \left( C_3^m - C_3^k - C_3^l \right)
+ \frac{4}{3} \left( S(S+1) - s_k (s_k+1) - s_l (s_l+1) \right)
\end{equation}
where $S$ denotes the total spin of the subsystem of the quark species $k$ and
$l$. From Fig.~\ref{fig:magnetic} one sees, that one can neglect the color
magnetic contribution for the charm quark. Then one can calculate the color
magnetic interaction for the light quarks for the different possible cases of
spin and color by taking care of the overall Fermi statistics.
For example, for three light quarks \{uds\} in a color octet state 
one gets $C_3^q = C_3^s = 16/3$, $C_3^c = 12$, $a_{qq} = -8$, $a_{qs} = -6$
or
$C_3^q = 40/3$, $C_3^s = 16/3$, $C_3^c = 12$, $a_{qq} = -4/3$, $a_{qs} =
-38/3$, respectively. Here $q$ stands for up- and down- quarks.
For the color singlet case, only one combination is possible with
$C_3^q = C_3^s = 16/3$, $C_3^c = 0$, $a_{qq} = -8$ which is just 
the case for the $\Lambda$ hyperon.
With the coefficients determined above, we calculated the masses of all
possible charmlet candidates within the modified bag model 
up to 6 light and 6 charm quarks. The result is shown in
Table~\ref{tab:bagcand} for charm quarks and in Table~\ref{tab:baganticand} for
anticharm quarks. Only in two case, indicated by a star, the threshold mass for
the strong decay is known experimentally. All the other threshold values are
estimated within the same bag calculation. As it is evident from the tables,
none of the candidates has a mass lower than the (estimated) threshold mass,
hence they are all unstable. 
The bag calculation might be accurate within 3\% guided by our fit to the
hadron spectra. 
There are four candidates which have a mass within
this range of 3\% above the threshold mass, three with baryon number two: 
\{cssudq\} with charge +1 or 0,
\{ccsudq\} with charge +2 or +1, \{ccssud\} with charge +1, and one with baryon
number three: \{cccssuudd\} with charge +2.
None of the anticharm candidates comes so close to the threshold. The reason
is that the decay products of the latter ones are D-mesons, which have a bigger
binding energy than $\Lambda_c$ baryons due to their smaller radius.

The dependence of the color magnetic and electric potential is quite sensitive
to the radius of the bag. A larger bag feels therefore a smaller attractive
potential than a smaller one. Secondly, the finite mass of the strange quark
also decreases the color magnetic interaction. These effects can then result in
an unbound candidate, which was bound in the flavor symmetric case.
Nevertheless, there is another more important term, which determines whether or
not a multi quark bag is bound. This is the so called Casimir energy which is
proportional to $-Z_0/R$ as also discussed in \cite{Zouzou94}. 
As the decay products will involve at least two bags,
this attractive term alone will raise the multi quark bag considerably.
In the case of equal radii, say 1 fm, 
one gets an additional shift of $Z_0/R\approx 300$ MeV! In reality, this shift
will be even higher, as the radius increases proportional to the number of
quarks sitting in the bag. 
Note, that this term is purely phenomenological and absorbs all effects
proportional to $1/R$ for the hadron spectra. This might change considerably
for multi quark states, but can not be determined within this simple bag model.
All the charmed candidates below a mass of 8 GeV are lying within 300 MeV above
the threshold. For the anticharm candidates, only the pentaquarks and the
tetraquarks fulfill this constraint.

Furthermore, the bag model assumes a spherical symmetric shape and an 
uniform radius for all quark species. This assumes, that the charm quarks are
sitting in the core surrounded by the light quarks.
As demonstrated in the previous sections,
the color electric term between the light and heavy quarks might be the
dominant one. This will cause the charmlet to form a highly nonspherical
configuration, as the light quarks are located around each charm quark, while
the charm quarks, feeling a repulsive potential between each other, are
separated from each other. This structure 
has to be studied in a fully three-dimensional model which is beyond the scope
of the present investigation.


\section{Summary}

We demonstrate that charm matter can be produced at the coming RHIC collider
in reasonable amounts. 
We estimate the production rates of these exotic new form of matter at RHIC
in a simple coalescence approach and find,
that about 10--100 multiple charmed exotics can be measured
per year. 
Energy loss of the charm quarks and hydrodynamical effects can enhance
these production rates considerably. 

We investigate the properties of charmed matter 
and finite multiple charmed objects.
The one-gluon exchange in bulk matter 
gets attractive for a
sufficiently high mass of the charm quark, much more than for the lighter 
strange quark. 
Charm matter is bound with respect to
strong decay to hadrons within an area of charm fraction of $f_c=1-3$ and a
strangeness fraction $f_s=0-1.5$ even if strange matter is unstable.

We introduce a modified bag model where the parameters are fitted to hadron
masses. 
The model can describe the masses of hadrons with and
without charm on the level of a few percent. The predictions of the masses for
multiply charm hadrons are a little bit higher than the ones from potential
models. 

We discuss then the color magnetic and color electric potentials for multi
quark states with charm. 
For finite systems, we show that in the limit of SU(3) flavor symmetry
the color magnetic interaction  can be more attractive than for the
corresponding hadron with a similar quark content. 
Systems with 3 and 4 light (u,d,s) quarks have an
enhanced color magnetic term compared to ordinary hadrons when adding heavy
quarks.
The constraint of color
neutrality can be released for the light quarks as the heavy charm quarks
neutralizes the overall quark bag. This allows for an additional freedom for
the spin and color structure of the light quarks, therefore enhancing the
binding energy. 
In addition, the color electric interaction between the charm
quarks is also attractive and increases with the number of charm quarks. 
It turns out that the dominant contribution to the binding of multi charm
objects comes from the color electric term between heavy and light quarks
which depends solely on the internal color structure.
These effects result in an enhanced stability of multiply charmed
exotics. 

For flavor symmetry and ignoring finite size effects, several charmlet
candidates have binding energy of more than 100 MeV.
A full calculation within the modified bag model does not give a bound
candidate. Also the pentaquark and the tetraquark are not bound in our
approach.  For some other cases the masses are lying just a few percent
(within our range of uncertainty, i.e.\ less than 3\%) 
above threshold. These candidates are:
the hexaquarks \{cssuud\}, \{cssudd\} with charge +1 and 0,
\{ccsuud\}, \{ccsuud\} with charge +2 or +1, 
\{ccssud\} with charge +1, and one candidate for $C=3$ 
\{cccssuudd\} with charge +2.
These candidates might be bound, if one replaces the charm quarks with the
heavier bottom quarks, as the attractive color electric forces will increase
for heavier quarks. Note, that the Tetraquark is bound in the limit of
$m_Q\to\infty$. We checked this for the above candidates but found that the
candidates with bottom quarks are still lying above the estimated threshold.
The main reason for the sudden 
instability of the candidates in the full bag model is,
that the effective energy correction $-Z_0/R$ shifts multi quark states up by
at least $+300$ MeV. Another reason is that non-spherical solutions might
exist which are lying lower in energy. This picture is supported by the
strongly 
attractive potential between light and heavy quarks which will cause the charm
quarks to be separated and not sitting in the center of the bag as assumed in
our calculation. This can only be addressed in full three-dimensional
calculation. Further investigations along this line using other models,
like potential models or lattice calculations, are certainly needed
to clarify this question. In this context, it is interesting to note that
meson exchange contributions are important for binding multiquark systems
as pointed out by several authors \cite{Mano93,Eric93,Torn94,Pepin97}. 

As the charmed hadrons live on a
timescale around $\tau\approx 100\mu$ a vertex detector is needed for their
detection which must be 
sitting extremely close to the target. Micro-strip detectors
are used for the CHARM200 and CERN COMPASS experiments which are looking for
pentaquark and tetraquark states in pp collisions. Similar techniques can be
used for searches at the RHIC collider.
Charm exotics as well as hadrons with two or three quarks can be well produced
at this energy. 
Hence,
truly heavy ion collisions at RHIC opens the unique and 
tantalizing opportunity of producing and detecting
multiply charm objects.
They can possibly be detected
by looking for a cascade of several weak decays
of the charm quarks.


\acknowledgments

We would like to thank Larry McLerran and Henning Heiselberg for
helpful discussions.
J.S.B. thanks Avraham Gal and the members of 
the Racah Institute of Physics at Hebrew University in Jerusalem,
Israel for their warm hospitality, where part of the calculations were done.
J.S.B. is supported 
by the Alexander-von-Humboldt Stiftung with a Feodor-Lynen
fellowship and in part by 
the Director, Office of Energy Research,
Office of High Energy and Nuclear Physics, Nuclear Physics Division of the
U.S. Department of Energy under Contract No.\ DE-AC03-76SF00098.



%
%

\begin{table}[htbp]
\caption{Hadron masses in the modified bag model (B$^{1/4}=0.1684$ GeV,
$\alpha_c=0.4764$, $Z_0=1.585$, $m_s=0.342$ GeV).}
\begin{tabular}{ccccc}
particle & quarks & $m_{\rm exp}$ [GeV] & $m_{\rm th}$ [GeV] & 
  $\Delta m/m_{\rm exp}$ \cr
\hline
N & qqq & 0.939 & 0.939 & fit \cr
$\rho$, $\omega$ & q\=q & 0.776 & 0.776 & fit \cr
$\Delta$ & qqq & 1.232 & 1.232 & fit \cr
$\pi$, $\eta$ & q\=q & 0.343 ? & 0.322 & --6.0\% ? \cr
$\Omega^-$ & sss & 1.673 & 1.673 & fit \cr
K & s\=q & 0.496 & 0.558 & +12.4\% \cr
K$^*$ & s\=q & 0.894 & 0.920 & +2.9\% \cr
$\Lambda$ & qqs & 1.116 & 1.111 & --0.4\% \cr
$\Sigma$ & qqs & 1.191 & 1.153 & --3.2\% \cr
$\Sigma^*$ & qqs & 1.384 & 1.381 & --0.2\% \cr
$\Xi$ & qss & 1.318 & 1.302 & --1.3\% \cr
$\Xi^*$ & qss & 1.533 & 1.528 & --0.3\% \cr
$\phi$ & s\=s & 1.019 & 1.060 & +4.0\% 
\end{tabular}
\label{tab:masses}
\end{table}

\begin{table}[htbp]
\caption{Hadron masses with charm ($m_c=1.788$ GeV).}
\begin{tabular}{ccccc}
particle & quarks & $m_{\rm exp}$ [GeV] & $m_{\rm th}$ [GeV] & 
  $\Delta m/m_{\rm exp}$ \cr
\hline
$\Lambda_c$ & qqc & 2.29 & 2.29 & fit \cr
D & c\=q & 1.87 & 1.82 & --2.8\% \cr
D$^*$ & c\=q & 2.01 & 2.01 & --0.1\% \cr
D$_s$ & c\=s & 1.97 & 1.98 & +0.7\% \cr
D$^*_s$ & c\=s & 2.11 & 2.14 & +1.3\% \cr
$\eta_c$ & c\=c & 2.98 & 3.05 & +2.4\% \cr
J/$\psi$ & c\=c & 3.10 & 3.15 & +1.9\% \cr
$\Sigma_c$ & qqc & 2.45 & 2.42 & --1.2\% \cr
$\Xi_c$ & qsc & 2.47 & 2.48 & +0.3\% \cr
$\Omega_c$ & ssc & 2.70 & 2.73 & +0.9\% \cr
$\Xi_{cc}$ & qcc & ? & 3.66 & \cr
$\Omega_{cc}$ & scc & ? & 3.82 & \cr
$\Omega_{ccc}$ & ccc & ? & 4.98 & \cr
$\Sigma^*_c$ & qqc & 2.53(?) & 2.53 & --0.1\% \cr
$\Xi^*_c$ & qsc & 2.64 & 2.67 & +1.0\% 
\end{tabular}
\label{tab:masscharm}
\end{table}

\begin{table}[htbp]
\caption{Hadron masses with beauty ($m_b=5.30$ GeV).}
\begin{tabular}{ccccc}
particle & quarks & $m_{\rm exp}$ [GeV] & $m_{\rm th}$ [GeV] & 
  $\Delta m/m_{\rm exp}$ \cr
\hline
$\Upsilon (1^-)$ & b\=b & 9.46 & 9.33 & --1.4\% \cr
B & b\=q & 5.28 & 5.21 & --1.2\% \cr
B$^*$ & b\=q & 5.32 & 5.31 & --0.2\% \cr
B$_s$ & b\=s & 5.37 & 5.37 & fit \cr
B$^*_s$ & b\=s & 5.42 & 5.45 & +0.5\% \cr
$\Lambda_b$ & qqb & 5.64 & 5.67 & +0.6\% \cr
$\eta_b$ & b\=b & ? & 9.26 & 
\end{tabular}
\label{tab:massbeauty}
\end{table}

\begin{table}[htbp]
\caption{
Charmed candidates and their main color magnetic and color electric 
interactions (q stands for u,d, and s quarks).}
\begin{tabular}{cccccccc}
Candidate & $V_M/M$ & $V'_E/E_{qc}$ & $V_E/E_{cc}$ &
decays to & $V_M/M$ & $V'_E/E_{qc}$ & $V_E/E_{cc}$ \cr
\hline
c5q [3] & --16 & --16/3 & 0 & cqq + qqq & --16 & --16/3 & 0 \cr
cc4q [$\bar 3$] & --16 & --16/3 & --8/3 & ccq + qqq & --8 & --16/3 & -8/3 \cr
cc4q [6] & --28/3 & --40/3 & +4/3 & ccq + qqq & --8 & --16/3 & -8/3 \cr
3c3q [1] & --8 & 0 & --8 & ccc + qqq & --8 & 0 & -8 \cr
3c3q [8] & --14 & --12 & --2 & ccc + qqq & --8 & 0 & -8 \cr
3c6q [1] & --24 & 0 & --8 & ccc + 2qqq & --16 & 0 & -8 \cr
3c6q [8] & --46/3 & --12 & --2 & ccc + 2qqq & --16 & 0 & --8 \cr
4c2q [$\bar 3$] & --8 & --16/3 & --8 & ccc + cqq & --8 & --16/3 & --8 \cr
4c2q [6] & --4/3 & --40/3 & --4 & ccc + cqq & --8 & --16/3 & --8 \cr
4c5q [3] & --16 & --16/3 & --8 & ccc+cqq+qqq & --16 & --16/3 & --8 \cr
4c5q [$\bar 6$] & --12 & --40/3 & --4 & ccc+cqq+qqq & --16 & --16/3 & --8 \cr
5cq [$\bar 3$] & 0 & --16/3 & --32/3 & ccc+ccq+qqq & 0 & --16/3 & --32/3 \cr
5c4q [$\bar 3$] & --16 & --16/3 & --32/3 & ccc+ccq+qqq & --8 & --16/3 & --32/3 \cr
6c3q [1] & --8 & 0 & --16 & 2ccc + qqq & --8 & 0 & --16 \cr
6c6q [1] & --24 & 0 & --16 & 2ccc + 2qqq & --16 & 0 & --16 \cr
6c [1] & 0 & 0 & --16 & 2ccc & 0 & 0 & --16 
\end{tabular}
\label{tab:cand}
\end{table}

\begin{table}[htbp]
\caption{
Anticharmed candidates and their main color magnetic and color electric 
interactions (q stands for u,d, and s quarks).}
\begin{tabular}{cccccccc}
Candidate & $V_M/M$ & $V'_E/E_{qc}$ & $V_E/E_{cc}$ &
decays to & $V_M/M$ & $V'_E/E_{qc}$ & $V_E/E_{cc}$ \cr
\hline
\=c4q [$\bar 3$] & --16 & --16/3 & 0 & D + qqq & --8 & --16/3 & 0 \cr
\=c\=cqq [$\bar 3$] & --8 & --16/3 & --8/3 & 2D & 0 & --32/3 & 0 \cr
\=c\=cqq [6] & --4/3 & --40/3 & +4/3 & 2D & 0 & --32/3 & 0 \cr
\=c\=c5q [$\bar 3$] & --16 & --16/3 & --8/3 & 2D + qqq & --8 & --32/3 & 0 \cr
\=c\=c5q [6] & --12 & --40/3 & +4/3 & 2D + qqq & --8 & --32/3 & 0 \cr
\=c\=c\=c3q [1] & --8 & 0 & --8 & \=c\=c\=c + qqq & --8 & 0 & --8 \cr
\=c\=c\=c3q [8] & --14 & --12 & --2 & \=c\=c\=c + qqq & --8 & 0 & --8 \cr
\=c\=c\=c6q [1] & --24 & 0 & --8 & \=c\=c\=c + 2qqq & --16 & 0 & --8 \cr
\=c\=c\=c6q [8] & --46/3 & --12 & --2 & \=c\=c\=c + 2qqq & --16 & 0 & --8 \cr
4\=cq [$\bar 3$] & 0 & --16/3 & --8 &  D + \=c\=c\=c & 0 & --16/3 & --8 \cr
4\=c4q [$\bar 3$] & --16 & --16/3 & --8 & D+\=c\=c\=c+qqq & --8 & --16/3 & --8 \cr
4\=c4q [6] & --28/3 & --40/3 & --4 & D+\=c\=c\=c+qqq & --8 & --16/3 & --8 \cr
5\=cqq [$\bar 3$] & --8 & --16/3 & --32/3 & 2D + \=c\=c\=c & 0 & --32/3 & --8 \cr
5\=c5q [$\bar 3$] & --16 & --16/3 & --32/3 &2D+\=c\=c\=c+qqq & --8 & --32/3 & --8\cr
6\=c3q [1] & --8 & 0 & --16 &2\=c\=c\=c + qqq & --8 & 0 & --16 \cr
6\=c6q [1] & --24 & 0 & --16 &2\=c\=c\=c + 2qqq & --16 & 0 & --16
\end{tabular}
\label{tab:anticand}
\end{table}

\begin{table}[htbp]
\caption{Estimate of the binding energy of charmed candidates
assuming $M = 18$ MeV,
$E_{qc} = 50$ MeV,
$E_{cc} = 70$ MeV (q stands for u,d, and s quarks).}
\begin{tabular}{ccccc}
Candidate & $\Delta V_M/M$ & $\Delta V'_E/E_{qc}$ & $\Delta V_E/E_{cc}$
& Energy gain [MeV] \cr
\hline
cc4q [$\bar 3$] & --8 & 0 & 0 & 144 \cr
cc4q [6] & --4/3 & --8 & +4 & 144 \cr
3c3q [8] & --6 & --12 & +6 & 288 \cr
3c6q [1] & --8 & 0 & 0 & 144 \cr
3c6q [8] & +2/3 & --12 & +6 & 168 \cr
4c2q [6] & +20/3 & --8 & +4 & 0 \cr
4c5q [$\bar 6$] & +4 & --8 & +4 & 48 \cr
5c4q [$\bar 3$] & --8 & 0 & 0 & 144 \cr
\hline
\=c4q [$\bar 3$] & --8 & 0 & 0 & 144 \cr
\=c\=cqq [$\bar 3$] & --8 & +16/3 & --8/3 & 64 \cr
\=c\=cqq [6] & --4/3 & --8/3 & +4/3 & 117 \cr
\=c\=c5q [$\bar 3$] & --8 & +16/3 & --8/3 & 64 \cr
\=c\=c5q [6] & --4 & --8/3 & +4/3 & 165 \cr
\=c\=c\=c3q [8] & --6 & --12 & +6 & 288 \cr
\=c\=c\=c6q [1] & --8 & 0 & 0 & 144 \cr
\=c\=c\=c6q [8] & +2/3 & --12 & +6 & 168 \cr
4\=c4q [$\bar 3$] & --8 & 0 & 0 & 144 \cr
4\=c4q [6] & --4/3 & -8 & +4 & 144 \cr
5\=cqq [$\bar 3$] & --8 & +16/3 & --8/3 & 64 \cr
5\=c5q [$\bar 3$] & --8 & +16/3 & --8/3 & 64 \cr
6\=c6q [1] & --8 & 0 & 0 & 144 
\end{tabular}
\label{tab:esti}
\end{table}

\begin{table}[htbp]
\caption{
Charmed candidates and their masses in the modified bag model
(q denotes the up or down quark, a star indicates that the threshold mass is
known).} 
\begin{tabular}{cccc}
candidate & charge & mass [GeV] & estimated threshold [GeV] \cr
\hline
csudud & +1 & 3.58 & 3.40$^*$ \cr
cssudq & 0,+1 & 3.66 & 3.58$^*$ \cr
ccsudq & +2,+1 & 4.87 & 4.76 \cr
ccssud & +1 & 5.04 & 4.94 \cr
cccsud & +2 & 6.29 & 6.10 \cr
cccssudud & +2 & 7.35 & 7.21 \cr
4cud & +3 & 7.53 & 7.27 \cr
4csq & +3,+2 & 7.72 & 7.45 \cr
4csudud & +3 & 8.70 & 8.38 \cr
4cssudq & +3,+2 & 8.84 & 8.56 \cr
5cq & +4,+3 & 9.00 & 8.64 \cr
5cs & +3 & 9.14 & 8.80 \cr
5csudq & +4,+3 & 10.0 & 9.76 \cr
5cssud & +3 & 10.2 & 9.92 \cr
6c & +4 & 10.3 & 9.96 \cr
6csud & +4 & 11.5 & 11.1 \cr
6cssudud & +4 & 12.6 & 12.2 
\end{tabular}
\label{tab:bagcand}
\end{table}

\begin{table}[htbp]
\caption{
Anticharmed candidates and their masses in the modified bag model
(q denotes the up or down quark).}
\begin{tabular}{cccc}
candidate & charge & mass [GeV] & threshold [GeV] \cr
\hline
\=csudq & 0,--1 & 3.13 & 2.91 \cr
\=cssud & --1 & 3.31 & 3.09 \cr
\=c\=cud & --1 & 4.04 & 3.73 \cr
\=c\=csq & --1,--2 & 4.23 & 3.84 \cr
\=c\=csudud & --1 & 5.24 & 4.78 \cr
\=c\=cssudq & --1,--2 & 5.38 & 4.88 \cr
4\=cq & --2,--3 & 7.25 & 6.85 \cr
4\=cs & --3 & 7.39 & 6.95 \cr
4\=csudq & --2,--3 & 8.33 & 7.57 \cr
4\=cssud & --3 & 8.51 & 7.67 \cr
5\=cud & --3 & 9.27 & 8.71 \cr
5\=csq & --3,--4 & 9.49 & 8.82 \cr
5\=csudud & --3 & 10.4 & 9.44 \cr
5\=cssudq & --3,--4 & 10.6 & 9.54 
\end{tabular}
\label{tab:baganticand}
\end{table}

%
%

\begin{thebibliography}{10}

\bibitem{Richard94}
J.-M. Richard,  in {\em {Proceedings of the International Conference on the
  future of high-sensitivity charm experiments}} (Fermilab, Batavia, 1994).

\bibitem{Mano93}
A.~V. Manohar and M.~B. Wise, Nucl. Phys. B {\bf 399},  17  (1993).

\bibitem{Eric93}
T.~E.~O. Ericson and G. Karl, Phys. Lett. B {\bf 309},  426  (1993).

\bibitem{Torn94}
N.~A. T\"ornqvist, Z. Phys. C {\bf 61},  525  (1994).

\bibitem{Carlson88}
J. Carlson, L. Heller, and J. Tjon, Phys. Rev. D {\bf 37},  744  (1988).

\bibitem{Pepin97}
S. Pepin, F. Stancu, M. Genovese, and J.-M. Richard, Phys. Lett. B {\bf 393},
  119  (1997).

\bibitem{Lipkin87}
H. Lipkin, Phys. Lett. B {\bf 195},  484  (1987).

\bibitem{Gig87}
C. Gignoux, B. Silvestre-Brac, and J.-M. Richard, Phys. Lett. B {\bf 193},  323
   (1987).

\bibitem{Chow95}
C.-K. Chow, Phys. Rev. D {\bf 51},  6327  (1995).

\bibitem{Moin96}
M. Moinester, Z. Phys. A {\bf 355},  349  (1996).

\bibitem{MALL96}
M. Moinester, D. Ashery, L. Landsberg, and H. Lipkin, Z. Phys. A {\bf 356},
  207  (1996).

\bibitem{Lin97}
Z. Lin, R. Vogt, and X.-N. Wang, Energy loss effects on charm and bottom
  production, nucl-th/9705006, 1997.

\bibitem{LZ93}
P. Levai and J. Zimanyi, Phys. Lett. B {\bf 304},  203  (1993).

\bibitem{Dover90}
C. Dover, Production of rare composite objects, preprint BNL-44520, 1990,
  presented at HIPAGS 1990.

\bibitem{FMNR94}
S. Frixione, M. Mangano, P. Nason, and G. Ridolfi, Nucl. Phys. B {\bf 431},
  453  (1994).

\bibitem{Vogt96}
R. Vogt, Z. Phys. C {\bf 71},  475  (1996).

\bibitem{SU97}
B. Svetitsky and A. Uziel, Phys. Rev. D {\bf 55},  2616  (1997).

\bibitem{Greiner91}
C. Greiner and H. St\"ocker, Phys. Rev. D {\bf 44},  3517  (1991).

\bibitem{Greiner88}
C. Greiner, D. Rischke, H. St\"ocker, and P. Koch, Phys. Rev. D {\bf 38},  2797
   (1988).

\bibitem{Spieles96}
C. Spieles {\it et~al.}, Phys. Rev. Lett. {\bf 76},  1776  (1996).

\bibitem{Chin79}
S. Chin and A. Kerman, Phys. Rev. Lett. {\bf 43},  1292  (1979).

\bibitem{Kettner95}
C. Kettner, F. Weber, M. Weigel, and N. Glendenning, Phys. Rev. D {\bf 51},
  1440  (1995).

\bibitem{DJJK75}
T. DeGrand, R. Jaffe, K. Johnson, and J. Kiskis, Phys. Rev. D {\bf 12},  2060
  (1975).

\bibitem{Izatt82}
D. Izatt, C. Detar, and M. Stephenson, Nucl. Phys. B {\bf 199},  269  (1982).

\bibitem{Free78}
B. Freedman and L. McLerran, Phys. Rev. D {\bf 17},  1109  (1978).

\bibitem{Hasen80}
P. Hasenfratz, R. Horgan, J. Kuti, and J. Richard, Phys. Lett. B {\bf 94},  401
   (1980).

\bibitem{SVZ79}
M. Shifman, A. Vainshtein, and V. Zakharov, Nucl. Phys. B {\bf 147},  448
  (1979).

\bibitem{Zouzou94}
S. Zouzou and J.-M. Richard, Few Body System {\bf 16},  1  (1994).

\bibitem{Don80}
J. Donoghue and K. Johnson, Phys. Rev. D {\bf 21},  1975  (1980).

\bibitem{Martin95}
A. Martin and J.-M. Richard, Phys. Lett. B {\bf 355},  345  (1995).

\bibitem{Jaffe91}
R. Jaffe, Nucl. Phys. B (Proc. Suppl.) {\bf 24B},  8  (1991).

\bibitem{HS78}
H. H\"ogaasen and P. Sorba, Nucl. Phys. B {\bf 145},  119  (1978).

\bibitem{Jaffe77}
R. Jaffe, Phys. Rev. Lett. {\bf 38},  195  (1977).

\bibitem{IN66}
C. Itzykson and M. Nauenberg, Rev. Mod. Phys. {\bf 38},  95  (1966).

\bibitem{Jaffe77b}
R. Jaffe, Phys. Rev. D {\bf 15},  281  (1977).

\end{thebibliography}
\end{document}